\documentclass[twocolumn,prl]{revtex4}

\usepackage{ORI_Group_style}
\usepackage[normalem]{ulem}
\usepackage{float}

\begin{document}

\title{Coherent Inflation for Large Quantum Superpositions of Microspheres}

\author{Oriol Romero-Isart}

\affiliation{Institute for Quantum Optics and Quantum Information of the
Austrian Academy of Sciences, A-6020 Innsbruck, Austria.}

\affiliation{Institute for Theoretical Physics, University of Innsbruck, A-6020 Innsbruck, Austria.}

\begin{abstract}

We show that coherent inflation, namely quantum dynamics generated by inverted conservative potentials acting on the center of mass of a massive object, is an enabling tool to prepare large spatial quantum superpositions in a double-slit experiment.  Combined with cryogenic, extreme high vacuum, and low-vibration environments, we argue that it is experimentally feasible to exploit coherent inflation  to prepare the center of mass of a micrometer-sized object in a spatial quantum superposition comparable to its size. In such a hitherto unexplored parameter regime gravitationally-induced decoherence could be unambiguously falsified. We present a protocol to implement coherent inflation in a double-slit experiment by letting a levitated microsphere traverse a static potential landscape. Such a protocol could be experimentally implemented with an all-magnetic scheme using superconducting microspheres.

\end{abstract}

\maketitle

%%%%%%%%%%%%%%%%%%%%%%%%%%%%%%%%%%%%%%%%%%%%%%%%%%%%%%%%%%%%%%%%%%%%%%%%%%%%%%%%%%%%%%%%%%%%

Delocalizing the center-of-mass of an object of mass $M$ over a distance $d$ is a fascinating possibility allowed by quantum mechanics. Such a quantum delocalization can be neatly observed in the double-slit experiment, where the interference pattern downstream shows that a quantum superposition of size $d$ (given by the slit separation) was prepared. It is fundamentally interesting to experimentally explore the parameter space $(M,d)$.  So far, impressive experiments in matter-wave interferometry explore the $(\text{small}~M, \text{large}~d)$ corner~\cite{Arndt1999,Cronin2009,Hornberger2012,Arndt2014}. In particular, quantum interference of molecules of  $M \sim 10^4$ atomic mass units (amu)~\cite{Eibenberger2013} as well as atom delocalization over half a meter~\cite{Kovachy2015} has been demonstrated. On the other hand, the recent field of cavity quantum micromechanics~\cite{revopt} explores the $(\text{large}~M, \text{small}~d)$ corner. Individual mechanical degrees of freedom in masses of the order of $M \sim 10^{13}$ amu have been cooled to the quantum ground state~\cite{oconell10,teufel11,chan11}. Such states have the  mass $M$ delocalized over tiny distances $d$ much smaller than a Bohr radius.

Levitated micromechanical oscillators~\cite{ORI2010, Chang2010, Barker2010,Gieseler2012,Kiesel2013,Millen2015} are ideal candidates to enter into the hitherto unexplored parameter regime of $(\text{large}~M, \text{large}~d)$ by combining techniques of quantum micromechanics and matter-wave interferometry~\cite{ORI2011,ORI2011b}; we call such hybrid schemes {\em quantum micromechanical interferometers}.
Quantum micromechanical interferometers have been so far discussed with optically manipulated dielectric nanospheres for earth-based room temperature experiments~\cite{ORI2011,ORI2011b,Asenbaum2013,Bateman2014}, as well as for space-based environments~\cite{MAQRO, Kaltenbaek2015}. In this Letter we aim at identifying, from a general point of view and without having {\em \`a priori} any experimental setup in mind, the limits in the $(\text{large}~M, \text{large}~d)$ parameter regime that could be achieved with quantum micromechanical interferometers. This will be done by considering unavoidable decoherence in challenging yet feasible environmental conditions.  As a by-product, this analysis will yield the necessary requirements to attain such limits. Apart from the expected requirements of (i) extreme high vacuum and (ii) cryogenic temperatures to minimize decoherence due to scattering of air molecules and interaction with black-body radiation~\cite{Joos2003,Schlosshauer2007,ORI2011b}, we shall propose that a key requirement is (iii) to equip quantum micromechanical interferometers with {\em coherent inflation} (CI), namely with the possibility to use conservative inverted harmonic potentials to exponentially speed-up the coherent dynamics needed to prepare and probe quantum superpositions in the $(\text{large}~M, \text{large}~d)$ parameter regime. We will conclude that with (iv) sufficient vibration isolation, quantum micromechanical interferometers with CI could explore the parameter regime of $(M \approx 10^{13}~\text{amu}, d \approx 1~\mu\text{m})$, that is, to delocalize an object as massive as current micromechanical oscillators~\cite{revopt} to distances comparable to its size. We will argue that an ideal candidate to meet the necessary requirements (i-iv) are magnetically levitated superconducting microspheres coupled to  quantum circuits~\cite{RomeroIsart2012, Pino2016}. Among other possibilities, such a parameter regime could unambiguously falsify the intriguing long-standing conjecture that gravity breaks down the quantum superposition principle at sufficiently large $(M,d)$~\cite{Diosi1984,Penrose1996}.

Let us start by recalling that while levitated micromechanical oscillators do not suffer clamping losses~\cite{ORI2010, Chang2010}, its center-of-mass decoheres due to position localization decoherence (PLD) whenever the environment interacts with its position (\eg~scattering of air molecules)~\cite{Joos2003,Schlosshauer2007}. The dynamics along a given axis (say $x$-axis) of the center-of-mass position of a sphere of mass $M$ in the presence of PLD can be modelled by the following master equation~\cite{Joos2003,Schlosshauer2007}:
$\bra{x} \dot \rho \ket{x'} = \bra{x} [\Hop,\hat \rho] \ket{x'}/(\im \hbar) - \Gamma(x-x') \bra{x} \hat \rho \ket{x'}  $, where $\xop \ket{x} = x \ket{x}$. The coherent dynamics are described by the Hamiltonian $\Hop=\pop^2/(2M) + V(\xop)$, where $V$ is a given potential and $\com{\xop}{\pop}=\im \hbar$. The second term in the master equation describes PLD, namely the exponential decay in time of spatial coherences $\bra{x} \hat \rho \ket{x'} $ (for $x\neq x'$) with a rate given by the decoherence function $\Gamma(x-x')$. The decoherence function can be conveniently approximated~\cite{Gallis1990} to $\Gamma(x)/\gamma=1- \exp[- \Lambda x^2/\gamma] $, where the decoherence rate $\gamma$ (with units of Hz) and the localization parameter $\Lambda$ (with units of $\text{Hz}/\text{m}^2$) are two parameters that model the source of decoherence. This function has two limits depending on the ratio between the saturation distance $\lambda \equiv \sqrt{\gamma/\Lambda} $ and the coherence length $\xi$ of the state $\hat \rho$, defined as $\bra{x} \hat \rho \ket{x'} \approx 0$ for $|x-x'| \gtrsim \xi$;  the short wavelength SW (long wavelength LW) corresponding to $\xi \gg \lambda$ ($\xi \ll \lambda$), such that  $ \Gamma \approx \gamma $ ($ \Gamma \approx \Lambda x^2 $).  In the LW limit the master equation can hence be written as $\dot \rho =  [\Hop,\hat \rho]/(\im \hbar) - \Lambda [\xop,[\xop,\hat \rho]] $. In decoherence processes due to scattering of particles~\cite{Joos2003,Schlosshauer2007}, $\gamma$ can be related to the scattering rate and $\lambda$ to the de Broglie wavelength of the scattered particles. The SW (LW) limit corresponds then to the case when single scattering events can (cannot) resolve distances smaller than the coherence length $\xi$. As shown below, since the de Broglie wavelength of air molecules is many orders of magnitude smaller than the wavelength of a thermal photon, PLD due to scattering of air molecules is typically in the SW limit whereas decoherence due to scattering, emission, and absorption of black-body photons is in the LW limit.

\begin{figure}[t]
\centering
\includegraphics[width=  \columnwidth]{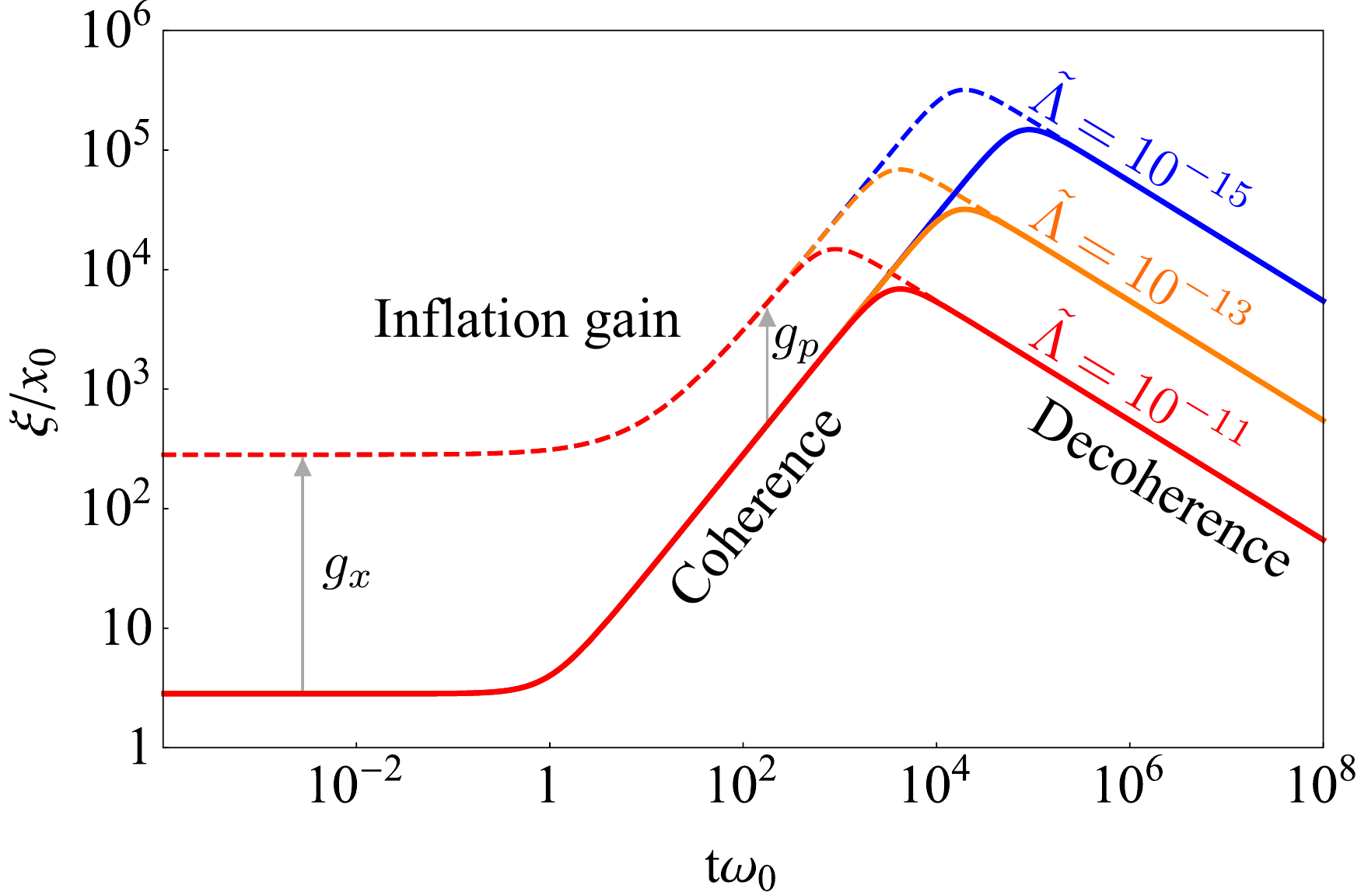}
\caption{ Coherence length $\xi$ in units of $x_0 \equiv \sqrt{v_x(0)}$ plotted as a function of time in units of $1/\w_0 \equiv 2 M v_x(0)/\hbar$. Solid lines show the evolution due to free expansion with PLD in the LW limit for different values of $\tilde \Lambda = \Lambda x_0^2/\w_0$, as indicated in the figure. The maximum coherence at  $t_\Lambda\w_0 =[3/(4 \tilde \Lambda)]^{1/3}$  is given by $\xi_\Lambda/x_0 =[32/(3 \tilde \Lambda^2)]^{1/6} $ . Dashed lines show the case in which CI with $g_x=500$ and $g_p=10$ has been previously performed. The maximum coherence length is then obtained at the shorter time $[3/(g_p^2 4 \tilde \Lambda)]^{1/3} \approx 0.21 t_\Lambda \w_0 $.
}
\label{Fig1}
\end{figure}

Let us now describe the first stage of a quantum micromechanical interferometer in the presence of PLD. Consider that at $t=0$ a levitated mechanical oscillator has been cooled to the ground state of the harmonic potential $V(\xop) = M \w_0^2 \xop^2/2$~\cite{WilsonRae07, Marquardt07, Genes2008, WilsonRae08}. Such a state belongs to the class of Gaussian states $\hat \rho$ with $\tr[\xop \hat \rho ] = \tr[\pop \hat \rho]=0$, and hence, it can be fully determined by the following three real parameters: $v_x \equiv \tr [ \xop^2 \hat \rho]$, $v_p \equiv \tr [ \pop^2 \hat \rho]$, and $c \equiv \tr [(\xop \pop + \pop \xop) \hat \rho]/2$. These parameters are constrained by the Heisenberg uncertainty principle: $v_x v_p - c^2 \geq \hbar^2/4$. The purity of a Gaussian state is given by $\mathcal{P}\equiv \tr [\hat \rho^2] =\hbar/(2 \sqrt{v_x v_p -c^2}) \leq 1$, and the coherence length $\xi$, defined by $\bra{x/2} \hat \rho \ket{-x/2} = \exp[-x^2/\xi^2]/\sqrt{2 \pi v_x}$, by $\xi=\mathcal{P} \sqrt{8 v_x}$.  At $t=0$, the ground state is pure with $v_x(0) = \hbar/(2M\w_0)$, $c(0)=0$, and with a tiny coherence length given by $\xi(0) = \sqrt{4\hbar/(M \w_0)}$. To be able to prepare large superpositions of the size $d$, it is convenient to grow $\xi$ such that $\xi \gtrsim d$. While in clamped mechanical oscillators this is very challenging~\cite{revopt}, levitated mechanical oscillators offer the possibility to switch off the trapping potential and let the system evolve freely with the Hamiltonian $\Hop= \pop^2/(2M)$~\cite{ORI2011,ORI2011b} . In the presence of PLD in the LW limit, the coherence length evolves as shown in \figref{Fig1}. Due to coherent dynamics, the coherence length grows linearly in time with a speed given by $\dot \xi \equiv \sqrt{8 v_p(0)}/M = \sqrt{4 \hbar \w_0/M}$ until decoherence starts to reduce the purity of the state and consequently the coherence length decays as $1/\sqrt{t}$. Due to decoherence in the LW limit, the coherence length has thus a maximum at time  $t_\Lambda = [3 M/(2 \hbar \Lambda \w_0 )]^{1/3}$ given by $\xi(t_\Lambda)=\sqrt{2/(\Lambda t_\Lambda)}$. Note that, although not shown in \figref{Fig1}, the growth of the coherence length is also limited to the timescale $1/\gamma$, where $\gamma$ is the decoherence rate associated to  any source of PLD in the SW limit (\eg~decoherence due to scattering of air molecules). For $t \gtrsim 1/\gamma$, $\xi$ decays exponentially in time. Therefore, the maximum coherence length will be given by $\xi^\star \equiv \xi( t^\star)$, where $t^\star(\Lambda,\gamma) \equiv \min \cpare{t_\Lambda,1/\gamma}$.

%%%%%%%%%%%%%%%%%%%%%%%%%%%%%%%%%%%%%%%%%%%%%%%%%%%%%%%%%%%%%%%%%%%%%%%%%%%%%%%%%%%%%%%%%%%%

At this point it is illustrative to discuss some numbers.  Consider a solid sphere of mass density $8570\,\text{Kg}/\text{m}^3$ (\eg~Niobium) that has been cooled to the ground state of an harmonic potential with trap frequency $\w_0 = 2\pi \times 10^5 \, \text{Hz}$.  The mass of the sphere is $M[\text{amu}] \approx 2 \times 10^{13} (R[\mu \text{m}])^3$ ($A[u]$ denotes the value of $A$ in $u$ units), where $R$ is its radius. At $t=0$ the coherence length is $\xi(0) [\text{m}] \approx 10^{-13} \times (R[\mu \text{m}])^{-3/2}$, which is much smaller than a Bohr radius for a sphere of one micrometer. In free coherent dynamics,  the coherence length grows linearly in time with a speed $\dot \xi [\text{nm}/\text{s}] \approx 86 / (R[\mu \text{m}])^{3/2}$. This is a relatively low speed since a sphere of one micrometer requires of the order of ten seconds to have a coherence length of the order of its radius, which poses a tremendous challenge in free fall experiments. Furthermore, there is unavoidable decoherence. On the one hand, thermal photons lead to PLD with~\cite{Joos2003,Schlosshauer2007} $\lambda= \pi^{2/3} \hbar c/(k_b T)$ and $\Lambda \approx 8!8 \zeta(9) cR^6 [k_b T/(\hbar c)]^9 \chi_R^2/(9 \pi)$ due to scattering and $\Lambda \approx 16 \pi^5 c R^3 [k_b T/(\hbar c)]^6 \chi_I/189$ due to emission and absorption (the bulk temperature of the sphere is assumed to be in equilibrium with the environmental temperature). Here $\chi_{R(I)}$ is the real (imaginary) part of $[\epsilon(\w_T)-1]/[\epsilon(\w_T) +2]$, where $\epsilon(\w_T)$ is the dielectric constant of the sphere at the thermal frequency $\w_T = k_b T/\hbar$. Since $\lambda[\text{mm}]\approx 5/T[\text{K}]$, decoherence due to thermal photons is well described in the LW limit. This leads to $t_\Lambda[\text{s}] \approx 7 \times 10^4 (T[\text{K}])^{-3}(R[\mu \text{m}])^{-1} \chi_R^{-2/3}$ for scattering and  $t_\Lambda[\text{s}] \approx 247 (T[\text{K}])^{-2} \chi_I^{-1/3}$ for emission and absorption of thermal photons. In room temperature experiments, as in optically manipulated dielectric spheres, coherent timescales of the orders of seconds are not available for spheres of one micrometer. Due to the strong dependence on the temperature, the situation is however very different in cryogenic environments (\eg~$T[\text{mK}]=100$), where decoherence due to black-body radiation can be safely neglected.  On the other hand, the sphere scatters air molecules of mass $m_a \approx 28.97\,\text{amu}$, temperature $T$, and pressure $P$. This leads to PLD with saturation distance given by~\cite{Joos2003, Schlosshauer2007} $\lambda= 2 \pi \hbar/\sqrt{2 \pi m_a k_b T}$ and $\gamma = 16 \pi \sqrt{2 \pi} P R^2/(\bar v m_a)$, where $\bar v$ is the thermal mean velocity of air molecules. One has that $\lambda[\text{nm}] \approx 0.32/\sqrt{T[\text{K}]}$ and $\gamma^{-1}[\text{s}] \approx 2 \times 10^{-16} \sqrt{T[\text{K}]} (P[\text{mbar}])^{-1} (R[\mu\text{m}])^{-2}$. In cryogenic temperatures and for coherence lengths larger than a nanometer, PLD due to scattering of air molecules is well described in the SW limit. For a sphere of a micrometer, extreme high vacuum of the order of $10^{-16}\,\text{mbar}$ is required to have coherent timescales $1/\gamma$ of the order of a second. Such vacuum levels are challenging but compatible with cryogenic environments~\cite{Gabrielse90,GabrielseRev}. In summary, in cryogenic environments $t^\star$ is given by the timescale $1/\gamma$ due to scattering of air molecules, which then leads to a maximum coherence length of the order of $\xi^\star[\text{nm}] \approx 2 \times 10^{-14} \sqrt{T[\text{K}]} (P[\text{mbar}])^{-1} (R[\mu\text{m}])^{-7/2} $.  This shows that even at extreme high vacuum in cryogenic environments, free dynamics for a sphere of one micrometer cannot grow its coherent length to scales comparable to its radius.  

%%%%%%%%%%%%%%%%%%%%%%%%%%%%%%%%%%%%%%%%%%%%%%%%%%%%%%%%%%%%%%%%%%%%%%%%%%%%%%%%%%%%%%%%%%%%

Let us now show how CI can improve this limitation. Instead of evolving the pure ground state of the harmonic trap (with $v_x(0)=\hbar/(2M \w_0)$ and $c(0)=0$) in free dynamics, consider that first one let the system evolve  for a short period of time $t_I$ in an inverted potential of frequency $\w_I$, namely $V(\xop)=- M \w_I^2 \xop^2/2$. Provided the evolution is purely coherent, one can show that the state is transformed to $v_x(t_I) = g_x^2 v_x(0)$,  $v_p(t_I) = g_p^2 v_p(0)$, and $2 c(t_I)=\hbar \sqrt{g^2-1}$, where the CI gain parameter $g \equiv g_x g_p$ is given by
\be
g =\sqrt{ 1+ \pare{\frac{\w_I^2+\w_0^2}{2 \w_I \w_0}}^2 \sinh^2(2 t_I \w_I)}.
\ee
For $g \gg1$ one has that $g_x/g_p \approx \w_0/\w_I$. Should the state evolve freely after CI, its coherence length $\xi$ would grow with a speed $g_p \approx \sqrt{\w_I g/\w_0}$ times larger, see \figref{Fig1}, and $\xi$ would reach its maximum due to decoherence in the LW limit at the shorter time $t_\Lambda /g_p^{2/3}$. 
The condition for the inflation to be purely coherent, namely that the purity is not reduced during $t_I$, requires
$\sinh(2 t_I \w_I) \ll M \w_I^3 \w_0 [\hbar \Lambda_I (\w_I^2 + \w_0^2)]^{-1}$. This
 upper bounds the CI gain to $g \ll g^\star$, where $g^\star \approx  \w_I^2 M/(2 \hbar \Lambda_I)$ (assuming $g^\star \gg 1$).
 The localization parameter $\Lambda_I$ during CI will be typically larger since externally applying a potential can easily lead to additional sources of decoherence. In particular, vibrations of the center of the potential lead to PLD decoherence in the LW limit with~\cite{Henkel1999,Schneider1999,Breuer2002} $\Lambda_v= M^2 \w_I^4 S_{xx}(\w_I)/(2 \hbar^2)$, where $S_{xx}(\w)$ is the power spectral density of the vibrations id the center of the potential (to be precise, PLD in the LW limit is obtained when $S_{xx}(\w_I) = S_{xx}(-\w_I)$, as in white noise). Considering that $\Lambda_I = \Lambda_v$, one has that the maximum inflationary gain is then given by $g^\star \approx \hbar/[M \w_I^2S_{xx}(\w_I)] $.  Using the same numbers as before and that the inverted harmonic potential has a frequency $\w_I [\text{Hz}]= 2 \pi \times 50$,  the maximum $g_p^\star = \sqrt{\w_I g^\star/\w_0}$ is given by $g_p^\star \approx 4 \times 10^{-15} /\sqrt{S_{xx}(\w_I)[\text{m}^2/\text{Hz}] (R[\mu\text{m}])^{3}}$. Hence,  useful CI ($g_p^\star \gg1$) for a sphere of one micrometer requires  $\sqrt{S_{xx} (\w_I)} \ll 10^{-15} \,\text{m}/\sqrt{\text{Hz}}$ in a cryogenic environment. As shown below, the vibration isolation required to use CI in the generation of fringes is more stringent. In a cryogenic environment with low vibration isolation, CI allows to grow the coherence length of a sphere of one micrometer  to a lengthscale comparable to its radius.

%%%%%%%%%%%%%%%%%%%%%%%%%%%%%%%%%%%%%%%%%%%%%%%%%%%%%%%%%%%%%%%%%%%%%%%%%%%%%%%%%%%%%%%%%%%%

After having generated a state with $\xi \sim R$, one could, in principle, prepare a large quantum superposition using a double slit with slit separation $d \lesssim \xi$ and slit width $\sigma \ll d$. Assuming that this crucial step can be done coherently, one would then obtain the state $\hat \rho_d = \ketbra{\psi}{\psi}$ with $\braket{x}{\psi} = \mathcal{N} \spare{\phi(x-d/2)+\phi(x+d/2)}$, where $\mathcal{N}$ is a normalization constant and $\phi(x) = \exp[-x^2/(4 \sigma^2)]$. A thorough analysis on how to prepare such a state with a continuous time quantum measurement of the $\xop^2$ observable can be found in~\cite{Pino2016}. Here we just focus on what happens  after having prepared $\hat \rho_d$. Since states after the double slit are non-Gaussian, it is more convenient to describe them via their Wigner function $W(x,p,t) = \intall \text{d} y \exp[- \im p y/\hbar] \bra{x+y/2} \hat \rho(t) \ket{x-y/2}/(2 \pi \hbar)$. The time evolution of the Wigner function in free dynamics with PLD in the LW limit is given by the differential equation $\partial_t W = - (p/M) \partial_x W + \hbar^2 \Lambda \partial^2_p W$. The position probability distribution $P(x,t) = \intall \text{d}p W(x,p,t)$ can then be expressed as
\be \label{eq:Blur}
P(x,t) = \intall \!P_{\Lambda = 0}(x+y,t) \frac{\exp \spare{-y^2/\sigma^2_\Lambda(t)}}{\sigma_\Lambda(t)\sqrt{\pi}} \text{d} y,
\ee
where $P_{\Lambda=0}(x,t)$ is the probability position distribution given in the absence of PLD, namely for $\Lambda=0$. PLD blurs the probability distribution with a blurring lengthscale given by $\sigma_{\Lambda}(t) = \sqrt{4\hbar^2 \Lambda t^3/(3 M^2)}$. One can show that  $P_{\Lambda=0}(x,t)$  exhibits fringes with a fringe separation given by $x_f(t) = 2 \pi \hbar t/(Md)$. Note that then $x_f/\sigma_\Lambda \sim t^{-1/2}$, and therefore, there is a time for which $\sigma_\Lambda(t) \gtrsim x_f(t)$ and fringes are erased due to PLD in the LW limit. Furthermore, recall that in the presence of PLD in the SW limit fringes would be exponentially erased at times larger than $1/\gamma$. The generation of visible fringes, which is required to demonstrate the preparation of a superposition state of separation $d$, puts a humongous limitation at large scales of mass and distance. This is so since the speed at which the fringes increase is given by $\dot x_f = 2 \pi \hbar/(Md)$, which for the numbers discussed above reads $\dot x_f [\text{nm}/\text{s}] \approx 2 \times 10^{-5} (d[\mu \text{m}])^{-1} (R[\mu \text{m}])^{-3} $. This means that for a sphere of one micrometer prepared in a superposition of one micrometer, $10^5$ seconds are required to generate fringes of the order of one nanometer. At this timescale fringes would be completely blurred due to scattering of air molecules even in extreme high vacuum.

%%%%%%%%%%%%%%%%%%%%%%%%%%%%%%%%%%%%%%%%%%%%%%%%%%%%%%%%%%%%%%%%%%%%%%%%%%%%%%%%%%%%%%%%%%%%

Let us now show how CI can crucially help to circumvent this limitation. In free expansion and without decoherence, the momentum operator at a given time $t_0$ is mapped into the position operator at a later time $t \gg t_0$ via $\xop(t) \approx \pop(t_0) M t$. With CI we aim at performing this mapping with an exponentially faster timescale. To this end, one has to do the following: let the system evolve, first, with the harmonic potential $V(\xop)= M \w_I^2 \xop^2/2$ for a time $t_R=\pi/(4 \w_I)$, and then, with the inverted potential $V(\xop)= -M \w_I^2 \xop^2/2$. One readily obtains that at time scales $t \gg t_R + 1/\w_I$, one has $\xop(t) \approx e^{\w_I t} \pop(t_0) /(\sqrt{2} M \w_I)$, as wanted.  Using the Wigner function formalism one can show~\cite{Pino2016} that $P(x,t)$ is also given by \eqnref{eq:Blur} where $P_{\Lambda=0}(x,t)$ exhibits, in this case, fringes with fringe separation $x_f(t) = \exp[\w_I t] 2 \pi \hbar/(M d \w_I)$. The blurring lengthscale with CI is given by~\cite{Pino2016}
\be
\sigma_\Lambda(t)= \sqrt{\frac{\hbar^2 \Lambda_I}{M^2 \w_I^3} \spare{\sinh(2 \w_I t) - 2 \w_I t}}.
\ee
Quite remarkably, note that for $\w_I t \gg1$ one has that  $x_f$ and $\sigma_\Lambda$ scale equally in time, as opposed to the free dynamics case. This means that during CI the inequality $x_f \gg \sigma_{\Lambda}$ will be time independent and thus fringes will not be erased due to PLD in the LW limit. CI allows to expand visible fringes in an exponentially faster time than in free evolution, and without blurring them, provided $\Lambda_I \ll  8 \pi^2 \w_I / d^2$ such that $x_f/\sigma_\Lambda \gg 1$ for $\w_I t \gg1$. This requires $\sqrt{S_{xx}(\w_I)} \ll 4 \pi \hbar/(d M \w_I^{3/2})$. Using the previous numbers, 
$\sqrt{S_{xx}(\w_I)} [\text{m}/\sqrt{\text{Hz}}] \ll 0.7 \times 10^{-17} (d[\mu \text{m}])^{-1} (R[\mu \text{m}])^{-3}$. For a sphere of one micrometer and a superposition of one micrometer, this is a very challenging but not impossible vibration isolation in a cryogenic environment. Indeed, a new generation of gravitational wave detectors in cryogenic environments (\eg~KAGRA project~\cite{kagra}) aims at vibration isolations many orders of magnitude better than what is required here.

%%%%%%%%%%%%%%%%%%%%%%%%%%%%%%%%%%%%%%%%%%%%%%%%%%%%%%%%%%%%%%%%%%%%%%%%%%%%%%%%%%%%%%%%%%%%

\begin{figure}[t]
\centering
\includegraphics[width=  \columnwidth]{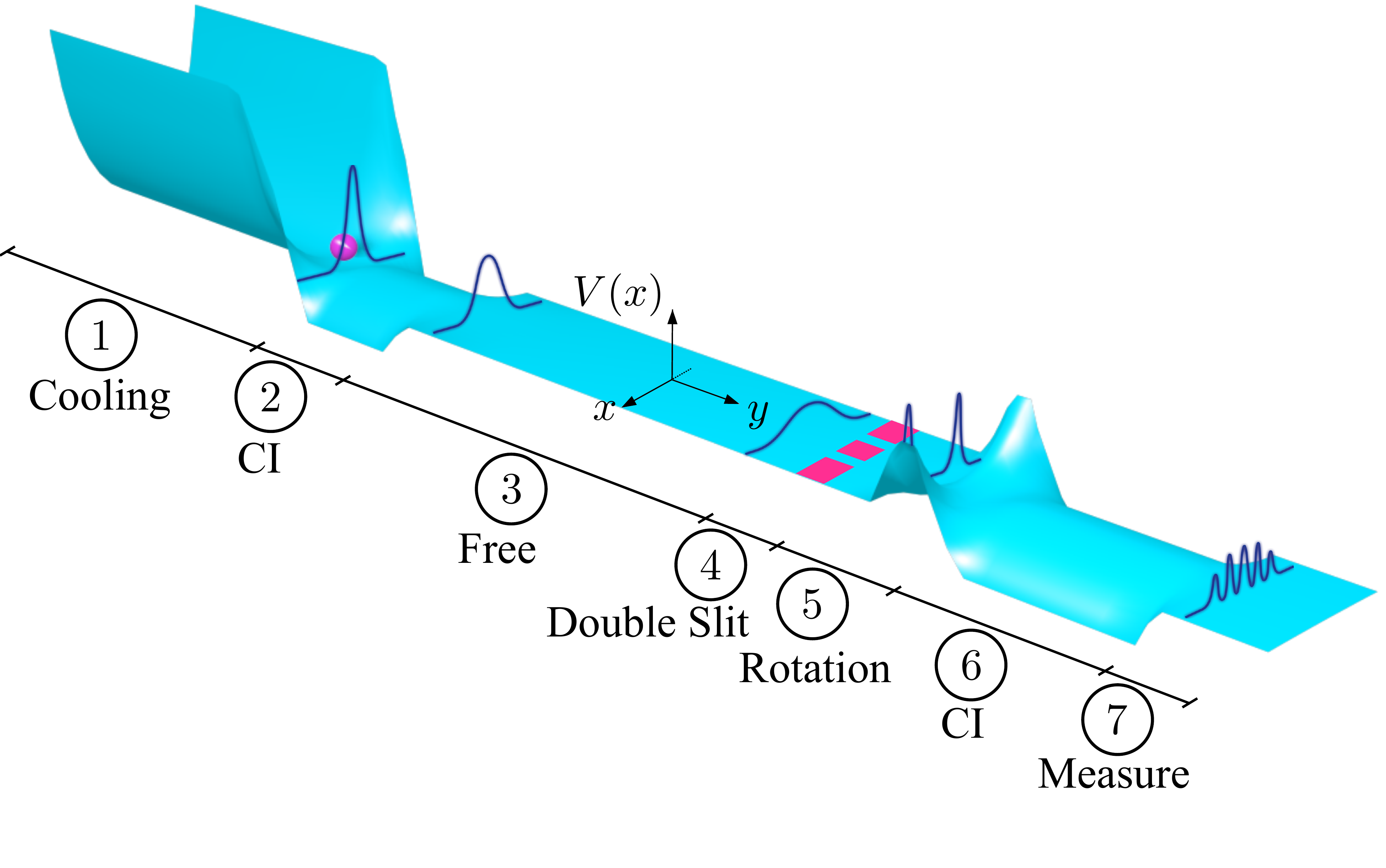}
\caption{ Quantum mechanical interferometer with CI. The potential for the $x$-coordinate is $V= M \w_0^2 \xop^2/2$ (step 1), $V= - M \w_I^2 \xop^2/2$ (step 2 \& 6), $V=0$ (step 3, 4 \& 7), and $V = M \w_I^2 \xop^2/2$ (step 5). Coupling to cavities in steps 1, 4, \& 7 are required to perform ground state cooling and quantum measurements. See~\cite{Pino2016} for details.  }
\label{Fig2}
\end{figure}

To implement CI in a double-slit experiment with minimal active control, we propose to let the sphere traverse, with a constant speed $v$ along the $y$-axis, the potential landscape $V(x)$ illustrated in~\figref{Fig2}. By properly engineering the potential landscape, the potential in the co-moving frame of the sphere would be time-dependent and would implement the quantum micromechanical interferometer with CI. By placing the potential landscape in the vicinity of cavities, ground state cooling, the continuous quantum measurement of $\xop^2$ performing the double slit, and the final position measurement to unveil the interference pattern, could also be passively implemented. As shown in this Letter, to implement such a protocol in the limit parameter regime of $(M \sim 10^{13}~\text{amu},d \sim 1~\mu\text{m})$, necessary requirements are: cryogenic temperatures ($T \lesssim 1~\text{K}$), both for the bulk of the sphere and its environment, extreme high vacuum ($P \approx 10^{-16}~\text{mbar}$),  extreme good vibration isolation ($\sqrt{S_{xx}} \approx 10^{-18}~\text{m}/\sqrt{\text{Hz}}$),  a conservative potential landscape, and coupling to quantum systems allowing the ground state cooling, the implementation of a double slit, and a final position measurement. All these requirements have been put together in a recent thorough proposal~\cite{Pino2016} consisting in a magnetically levitated superconducting microsphere traversing a static magnetic potential (so-called  {\em magnetic skatepark}) on top of a superconducting chip hosting flux-dependent quantum circuits.

%%%%%%%%%%%%%%%%%%%%%%%%%%%%%%%%%%%%%%%%%%%%%%%%%%%%%%%%%%%%%%%%%%%%%%%%%%%%%%%%%%%%%%%%%%%%

To conclude, we remark that the parameter regime $(M \sim 10^{13}~\text{amu},d \sim 1~\mu\text{m})$ opens many interesting possibilities, in particular with regard to gravity. For instance, the long-standing conjecture that gravity could cause the breakdown of the superposition principle at sufficiently large scales~\cite{Diosi1984,Penrose1996} can be formulated, for a sphere of radius $R$ and mass $M$, in terms of PLD with $\Lambda_G = GM^2/(2 \hbar R^3)$ and $\lambda_{G} = R$~\cite{ORI2011b}, where $G$ is the gravitational constant. This is the parameter-free prediction of gravitationally-induced decoherence, which  is derived from an homogeneous mass density, and leads to lower-bounded decoherence rates~\cite{Diosi2007}. For coherence lengths $\xi \lesssim R$, one can use the LW limit to obtain $t_\Lambda[\text{s}] \approx 1.26$ (for any $R$), which can be further reduced using CI. In accordance with our previous analysis, such a timescale makes it possible to falsify gravitationally-induced decoherence using a slit separation in the regime $\xi(\Lambda + \Lambda_G,\gamma) \ll d \ll \xi(\Lambda,\gamma)$, where $\Lambda$ and $\gamma$ are standard PLD sources, and $\xi(\Lambda,\gamma)$ is the coherence length just before the double slit. For further details see~\cite{Pino2016}.

%%%%%%%%%%%%%%%%%%%%%%%%%%%%%%%%%%%%%%%%%%%%%%%%%%%%%%%%%%%%%%%%%%%%%%%%%%%%%%%%%%%%%%%%%%%%

This work is supported by the European Research Council (ERC-2013-StG 335489 QSuperMag) and the Austrian Federal Ministry of Science, Research, and Economy (BMWFW). ORI acknowledges the contribution of H. Pino, J. Prat-Camps, K. Sinha, and B. P. Venkatesh.

 %%%%%%%%%%%%%%%%%%%%%%%%%%%%%%%%%%%%%%%%%%%%%%%%%%%%%%%%%%%%%%%%%%%%%%%%%%%%%%%%%%%%%%%%%%%%

%%%%%%%%%%%%%%%%%%%%%%%%%%%%%%%%%%%%%%%%%%%%%%%%%%%%%%%%%%%%%%%%%%%%%%%%%%%%%%%%%%%%%%%%%%%%

%%%%%%%%%%%%%%%%%%%%%%%%%%%%%%%%%%%%%%%%%%%%%%%%%%%%%%%%%%%%%%%%%%%%%%%%%%%%%%%%%%%%%%%%%%%%

%%%%%%%%%%%%%%%%%%%%%%%%%%%%%%%%%%%%%%%%%%%%%%%%%%%%%%%%%%%%%%%%%%%%%%%%%%%%%%%%%%%%%%%%%%%%

 %%%%%%%%%%%%%%%%%%%%%%%%%%%%%%%%%%%%%%%%%%%%%%%%%%%%%%%%%%%%%%%%%%%%%%%%%%%%%%%%%%%%%%%%%%%%

 %%%%%%%%%%%%%%%%%%%%%%%%%%%%%%%%%%%%%%%%%%%%%%%%%%%%%%%%%%%%%%%%%%%%%%%%%%%%%%%%%%%%%%%%%%%% 

%%%%%%%%%%%%%%%%%%%%%%%%%%%%%%%%%%%%%%%%%%%%%%%%%%%%%%%%%%%%%%%%%%%%%%%%%%%%%%%%%%%%%%%%%%%%


\begin{thebibliography}{27}

%\newcommand{\PRL}[3]{Phys.~Rev. Lett.~\textbf{#1}, #2~(#3)}
%\newcommand{\APL}[3]{Appl.~Phys. Lett.~\textbf{#1}, #2~(#3)}
%\newcommand{\EPL}[3]{Europhys.~Lett.~\textbf{#1}, #2~(#3)}
%\newcommand{\PRA}[3]{Phys.~Rev. A~\textbf{#1}, #2~(#3)}
%\newcommand{\PRB}[3]{Phys.~Rev. B~\textbf{#1}, #2~(#3)}
%\newcommand{\RMP}[3]{Rev.~Mod.~Phys.~\textbf{#1}, #2~(#3)}
%\newcommand{\PRD}[3]{Phys.~Rev. D~\textbf{#1}, #2~(#3)}
%\newcommand{\JPA}[3]{J.~Phys. A~\textbf{#1}, #2~(#3)}
%\newcommand{\PLA}[3]{Phys.~Lett. A~\textbf{#1}, #2~(#3)}
%\newcommand{\JOB}[3]{J.~Opt. B~\textbf{#1}, #2~(#3)}
%\newcommand{\JMP}[3]{J.~Math.~Phys.~\textbf{#1}, #2~(#3)}
%\newcommand{\JMO}[3]{J.~Mod.~Opt.~\textbf{#1}, #2~(#3)}
%\newcommand{\NAT}[3]{Nature~\textbf{#1}, #2~(#3)}
%\newcommand{\NATP}[3]{Nature Phys.~\textbf{#1}, #2~(#3)}
%\newcommand{\NATO}[3]{Nature Photon.~\textbf{#1}, #2~(#3)}
%\newcommand{\SCI}[3]{Science~\textbf{#1}, #2~(#3)}
%\newcommand{\IBID}[3]{{\em ibid}.~\textbf{#1}, #2~(#3)}
%\newcommand{\NJP}[3]{New~J.~Phys.~\textbf{#1}, #2~(#3)}
%\newcommand{\PNAS}[3]{Proc.~Natl.~Acad.~Sci.~U.S.A.~\textbf{#1}, #2~(#3)}

\expandafter\ifx\csname natexlab\endcsname\relax\def\natexlab#1{#1}\fi
\expandafter\ifx\csname bibnamefont\endcsname\relax
  \def\bibnamefont#1{#1}\fi
\expandafter\ifx\csname bibfnamefont\endcsname\relax
  \def\bibfnamefont#1{#1}\fi
\expandafter\ifx\csname citenamefont\endcsname\relax
  \def\citenamefont#1{#1}\fi
\expandafter\ifx\csname url\endcsname\relax
  \def\url#1{\texttt{#1}}\fi
\expandafter\ifx\csname urlprefix\endcsname\relax\def\urlprefix{URL }\fi
\providecommand{\bibinfo}[2]{#2}
\providecommand{\eprint}[2][]{\url{#2}}


\bibitem{Arndt1999}
M.~Arndt, O.~Nairz, J.~Vos-Andreae, C.~Keller, G.~van~der~Zouw, and A.~Zeilinger, \href{http://www.nature.com/nature/journal/v401/n6754/abs/401680a0.html}{Nature~(London) {\bf 401}, 680 (1999).}

\bibitem{Cronin2009}
A.~D.~Cronin, J.~Schmiedmayer, and D.~E.~Pritchard, \href{http://journals.aps.org/rmp/abstract/10.1103/RevModPhys.81.1051}{Rev.~Mod.~Phys. {\bf 81}, 1051 (2009).}

\bibitem{Hornberger2012}
K.~Hornberger, S.~Gerlich, P.~Haslinger, S.~Nimmrichter, and M.~Arndt, \href{http://journals.aps.org/rmp/abstract/10.1103/RevModPhys.84.157}{Rev.~Mod.~Phys. {\bf 84}, 157 (2012).}

\bibitem{Arndt2014}
M.~Arndt and K.~Hornberger, \href{http://www.nature.com/nphys/journal/v10/n4/full/nphys2863.html}{Nat.~Phys. {\bf 10}, 271 (2014).}

\bibitem{Eibenberger2013}
S.~Eibenberger, S.~Gerlich, M.~Arndt, M.~Mayor, and J.~T\"uxen, \href{http://pubs.rsc.org/en/Content/ArticleLanding/2013/CP/c3cp51500a#!divAbstract}{Phys.~Chem.~Chem.~Phys. {\bf 15}, 14696 (2013).}

\bibitem{Kovachy2015}
T.~Kovachy, P.~Asenbaum, C.~Overstreet, C.~A.~Donnelly, S.~M.~Dickerson,	A.~Sugarbaker,  J.~M.~Hogan, and M.~A.~Kasevich, \href{http://www.nature.com/nature/journal/v528/n7583/abs/nature16155.html}{Nature {\bf 528}, 530 (2015).}

\bibitem{revopt}
M.~Aspelmeyer, T.~J.~Kippenberg, and F.~Marquardt, \href{http://journals.aps.org/rmp/abstract/10.1103/RevModPhys.86.1391}{Rev.~Mod.~Phys {\bf 86}, 1391 (2014).}

\bibitem{oconell10}
A.~D.~O'Connell, M.~Hofheinz, M.~Ansmann, R.~C.~Bialczak, M.~Lenander, E.~Lucero, M.~Neeley, D.~Sank, H.~Wang, M.~Weides, J.~Wenner, J.~M.~Martinis, and A.~N.~Cleland, \href{http://www.nature.com/nature/journal/v464/n7289/abs/nature08967.html}{Nature~(London) {\bf 464}, 697 (2010).}

\bibitem{teufel11}
J.~D.~Teufel,	T.~Donner,	D.~Li,	J.~W.~Harlow,	M.~S.~Allman,	K.~Cicak,	A.~J.~Sirois,	J.~D.~Whittaker,	K.~W.~Lehnert, and R.~W.~Simmonds, \href{http://www.nature.com/nature/journal/v475/n7356/full/nature10261.html}{Nature~(London) {\bf 475}, 359 (2011).}

\bibitem{chan11}
J.~Chan,	T.~P.~Mayer~Alegre,	A.~H.~Safavi-Naeini,	J.~T.~Hill,	A.~Krause,	S.~Gr\"oblacher,	M.~Aspelmeyer, and O.~Painter, \href{http://www.nature.com/nature/journal/v478/n7367/abs/nature10461.html}{Nature~(London) {\bf 478}, 89 (2011).}

\bibitem{ORI2010}
O.~Romero-Isart, M.~L.~Juan, R.~Quidant, and J.~I.~Cirac, \href{http://iopscience.iop.org/article/10.1088/1367-2630/12/3/033015/meta}{New~J.~Phys. {\bf 12}, 033015 (2010).}

\bibitem{Chang2010}
D.~E.~Chang, C.~A.~Regal, S.~B.~Papp, D.~J.~Wilson, J.~Ye, O.~Painter, H.~J.~Kimble, and P.~Zoller, \href{http://www.pnas.org/content/107/3/1005}{Proc.~Nat.~Acad.~Sci.~U.~S.~A. {\bf 107}, 1005 (2010).}

\bibitem{Barker2010}
P.~F.~Barker and M.~N.~Schneider, \href{http://journals.aps.org/pra/abstract/10.1103/PhysRevA.81.023826}{Phys.~Rev.~A {\bf 81}, 023826 (2010).}

\bibitem{Gieseler2012}
J.~Gieseler, B.~Deutsch, R.~Quidant, and L.~Novotny, \href{http://journals.aps.org/prl/abstract/10.1103/PhysRevLett.109.103603}{Phys.~Rev.~Lett. {\bf 109}, 103603 (2012).}

\bibitem{Kiesel2013}
N.~Kiesel, F.~Blaser, U.~Delić, D.~Grass, R.~Kaltenbaek, and M.~Aspelmeyer, \href{http://www.pnas.org/content/110/35/14180.abstract}{Proc.~Nat.~Acad.~Sci.~U.~S.~A. {\bf 110}, 14180 (2013).}

\bibitem{Millen2015}
J.~Millen, P.~Z.~G.~Fonseca, T.~Mavrogordatos, T.~S.~Monteiro, and P.~F.~Barker, \href{http://journals.aps.org/prl/abstract/10.1103/PhysRevLett.114.123602}{Phys.~Rev.~Lett. {\bf 114}, 123602 (2015).}

\bibitem{ORI2011}
O.~Romero-Isart, A.~C.~Pflanzer, F.~Blaser, R.~Kaltenbaek, N.~Kiesel, M.~Aspelmeyer, and J.~I.~Cirac, \href{http://journals.aps.org/prl/abstract/10.1103/PhysRevLett.107.020405}{Phys.~Rev.~Lett. {\bf 107}, 020405 (2011).}

\bibitem{ORI2011b}
O.~Romero-Isart, \href{http://journals.aps.org/pra/abstract/10.1103/PhysRevA.84.052121}{Phys.~Rev.~A {\bf 84}, 052121 (2011).}

\bibitem{Asenbaum2013}
P.~Asenbaum,	S.~Kuhn, S.~Nimmrichter,	U.~Sezer,	and M.~Arndt, \href{http://www.nature.com/ncomms/2013/131106/ncomms3743/full/ncomms3743.html}{Nat.~Commun. {\bf 4}, 2743 (2013).}

\bibitem{Bateman2014}
J.~Bateman,	S.~Nimmrichter,	K.~Hornberger, and H.~Ulbricht, \href{http://www.nature.com/ncomms/2014/140902/ncomms5788/full/ncomms5788.html}{Nat.~Commun. {\bf 5}, 4788 (2014).}

\bibitem{MAQRO}
R.~Kaltenbaek, G.~Hechenblaikner, N.~Kiesel, O.~Romero-Isart, K.~C.~Schwab, U.~Johann, and M.~Aspelmeyer, \href{http://link.springer.com/article/10.1007\%2Fs10686-012-9292-3}{Exp.~Astron. \textbf{34}, 123~(2012).}

\bibitem{Kaltenbaek2015}
R.~Kaltenbaek \etal \href{http://arxiv.org/abs/1503.02640}{arXiv:1503.02640.}

\bibitem{Joos2003}
E.~Joos, H.~D.~Zeh, C.~Kiefer, D.~Giulini, J.~Kupsch, and I.~-O.~Smatescu, {\em Decoherence and the Appearance of a Classical World in Quantum Theory} (Springer, New York, 2003).

\bibitem{Schlosshauer2007}
M.~A.~Schlosshauer, {\em Decoherence and the Quantum-to-Classical Transition} (Springer, Berlin, 2007).

\bibitem{RomeroIsart2012}
O.~Romero-Isart, L.~Clemente, C.~Navau, A.~Sanchez, and J.~I.~Cirac, \href{http://journals.aps.org/prl/abstract/10.1103/PhysRevLett.109.147205}{Phys.~Rev.~Lett. {\bf 109}, 147205 (2012).}

\bibitem{Pino2016}
H.~Pino, J.~Prat-Camps, K.~Sinha, B.~P.~Venkatesh, and O.~Romero-Isart, \href{http://arxiv.org/abs/1603.01553}{arXiv:1603.01553}.

\bibitem{Diosi1984}
L.~Di\'{o}si, \href{http://www.sciencedirect.com/science/article/pii/0375960184903979}{Phys.~Lett.~A {\bf 105}, 199 (1984).}

\bibitem{Penrose1996}
R.~Penrose, \href{http://link.springer.com/article/10.1007/BF02105068}{Gen.~Relativ.~Gravit.~\textbf{28}, 581 (1996).}



\bibitem{Gallis1990}
M.~R.~Gallis and G.~N.~Fleming, \href{http://journals.aps.org/pra/abstract/10.1103/PhysRevA.42.38}{\PRA{42}{38}{1990}}.

\bibitem{WilsonRae07}
I.~ Wilson-Rae, N.~ Nooshi, W.~ Zwerger, and T.~J.~Kippenberg, \href{http://journals.aps.org/prl/abstract/10.1103/PhysRevLett.99.093901}{Phys.~Rev.~Lett. {\bf 99}, 093901 (2007).}

\bibitem{Marquardt07}
F.~Marquardt, J.~P.~Chen, A.~A.~Clerk, and S.~M.~Girvin, \href{http://journals.aps.org/prl/abstract/10.1103/PhysRevLett.99.093902}{Phys.~Rev.~Lett. {\bf 99}, 093902 (2007).}

\bibitem{Genes2008}
C.~Genes, D.~Vitali, P.~Tombesi, S.~Gigan, and M.~Aspelmeyer, \href{http://journals.aps.org/pra/abstract/10.1103/PhysRevA.77.033804}{Phys.~Rev.~A {\bf 77}, 033804 (2008).}

\bibitem{WilsonRae08}
I.~Wilson-Rae, N.~Nooshi, J.~Dobrindt, T.~J.~Kippenberg, and W.~Zwerger, \href{http://iopscience.iop.org/article/10.1088/1367-2630/10/9/095007/meta}{New.~J.~Phys. {\bf 10}, 095007 (2008).}

\bibitem{Gabrielse90} G.~Gabrielse, X.~Fei,  L.~A.~Orozco, R.~L.~Tjoelker,  J.~Haas, H.~Kalinowsky, T.~A.~Trainor, and W.~Kells, \href{http://journals.aps.org/prl/abstract/10.1103/PhysRevLett.65.1317}{Phys.~Rev.~Lett. {\bf 65}, 1317 (1990).}

\bibitem{GabrielseRev} G.~Gabrielse, \href{http://www.sciencedirect.com/science/article/pii/S1049250X01800379}{Advances in Atomic, Molecular, and Optical Physics \textbf{45}, 1 (2001).}


\bibitem{Henkel1999}
C.~Henkel, S.~P\"otting, M.~Wilkens, \href{http://link.springer.com/article/10.1007/s003400050823}{App.~Phys.~B {\bf 69}, 379 (1999).}

\bibitem{Schneider1999}
S.~Schneider and G.~J.~Milburn, \href{http://journals.aps.org/pra/abstract/10.1103/PhysRevA.59.3766}{\PRA{59}{3766}{1999}}.

\bibitem{Breuer2002}
H.~-P.~Breuer and F.~Petruccione, {\em The Theory of Open Quantum Systems} (Oxford University Press, Oxford, 2002).

\bibitem{kagra}
C.~Tokoku \etal \href{http://scitation.aip.org/content/aip/proceeding/aipcp/10.1063/1.4860850}{AIP~Conf.~Proc. {\bf 1573}, 1254 (2014).}

\bibitem{Diosi2007}
L.~Di\'osi, \href{http://iopscience.iop.org/article/10.1088/1751-8113/40/12/S07/meta;jsessionid=6A3176BC7BF382793F5870BCAAEF0DA7.c4.iopscience.cld.iop.org}{J.~Phys.~A:~Math.~Theor. {\bf 40}, 2989 (2007)}.


\end{thebibliography}
\end{document}